\documentclass[prx,twocolumn,amsmath,amssymb]{revtex4}
\usepackage[dvipdfmx]{xcolor}
\usepackage{bm}
\usepackage{dcolumn}
\usepackage[pdftex]{graphicx}    

\usepackage{slashed}
\usepackage{amsmath}
\usepackage{accents}
\usepackage{mathdots}
\usepackage{ulem}
\usepackage{physics}

\newcommand{\cin}[1]{{\color{black}#1}}

\begin{document}
	
	\title{
		Chern Numbers Associated with the Periodic Toda Lattice
	}
	
	\author{Kyoka Sato and Takahiro Fukui}
	\affiliation{Department of Physics, Ibaraki University, Mito 310-8512, Japan}

	\date{\today}
	
	\begin{abstract}
		The periodic Toda lattice is solved by exploiting the spectral properties of the Lax operator, in which the boundary states play an important role. 
		We show that these boundary states have a topological origin similar to that of the edge states in topological insulators,
		and consequently,
		that the bulk wave functions of the Lax operator yield nontrivial Chern numbers.
		This implies that the periodic Toda lattice belongs to the same topological class as the Thouless pump.
		We demonstrate that the cnoidal wave of the Toda lattice exhibits a Chern number of $-1$ per period.
	\end{abstract}
	
	\pacs{
	}
	
	\maketitle
	
	The Toda lattice \cite{Toda:1967aa, Toda:1967ab} is a one-dimensional model with nonlinear interaction. 
	The integrability of the model reveals various physical properties of the nonlinear wave propagation based on exact solutions.
	For example, the Toda lattice allows solitons on a lattice,
	\cin{and a} periodic solution, known as a cnoidal wave, can be interpreted as a sequence of solitons \cite{Toda:1967aa, Toda:1967ab}. 
	The inverse scattering method based on the Lax representation \cite{https://doi.org/10.1002/cpa.3160210503} 
	was developed on the lattice, 
	which enables us to obtain an $N$ soliton solution \cite{10.1143/PTP.51.703}.
	The Lax representation 
	is also useful for solving the periodic Toda lattice.
	Exploiting the spectral properties of the Lax operator, 
	the exact solution was explicitly written by
	the Riemann theta function \cite{Kac1975,Date:1976aa}.
	Here, the boundary states
	of the Lax operator play a crucial role. 
	That is, the solutions of the Toda lattice under \cin{the periodic boundary condition} are expressed using the spectrum of the Lax operator under the open boundary condition.

	In this letter, we shed light on the topological property of these boundary states in
	the periodic Toda lattice. Rather than being topological, the solitons of the Toda lattice are known to result from the balance between the nonlinear and dispersive effects. 
	The solitons of integrable systems are often referred to as non-topological solitons.
	Nevertheless, we demonstrate that the boundary states of the Lax operator have a topological origin similar to the edge states of topological 
	insulators.
	This argument is based on the theory of the bulk--edge correspondence in the quantum Hall effect \cite{Hatsugai:1993fk,Hatsugai:1993aa},
	which claims that the edge states are guaranteed by the bulk topological property.
	We show that the periodic Toda lattice is classified as topological class A \cite{Altland:1997aa,Zirnbauer:1996wi,Schnyder:2008aa}, 
	the same class to which the quantum Hall \cin{effect} in two dimensions \cite{Thouless:1982uq,kohmoto:85}
	and the Thouless pump in $1+1$ dimensions belong \cite{Thouless:1983fk,Nakajima:2016aa,Lohse:2016aa}.
	Thus, the solutions of the periodic Toda lattice can be characterized by the Chern numbers.

	Solitons are stable against small perturbations \cite{doi:10.1137/0131013}. 
	Naturally, topological solitons are protected topologically. 
	In this regard, it is necessary to point out that the stability of non-topological solitons in integrable systems
	also has a topological origin, 
	although the topology of non-topological solitons has not yet been considered. 
	The results of this study are valid only for the periodic solution of the Toda lattice, and do not apply directly to the soliton solutions for infinite chains. However, the periodic solution is a so-called soliton train \cite{Toda:1967ab}, 
	such that the topological stability of the periodic solution is relevant to 
	that of its component, the soliton itself.

	The Toda potential is defined as
	\begin{alignat}1
	\phi(x)&=\frac{a}{b}(e^{-bx}-1)+ax.
	\label{TodP}
	\end{alignat}
	The harmonic potential limit is $b\rightarrow0$ while keeping $ab$ finite, whereas the hard-core potential limit is $b\rightarrow \infty$.
	Let us consider the periodic system of $q$ particles described by the
	equations of motion (EOM) 
	$m\ddot x_j
	=\phi'(x_{j+1}-x_j)-\phi'(x_j-x_{j-1})
	$\cin{,}
	where $j=1,2,\cdots, q$ and $x_{j+q}=x_j$.
	In preparation, let us review the way in which to rewrite the EOM such that \cin{they are} suitable for the exact solution.
	\cin{First, we} introduce the dimensionless variables as follows:
	$Q_j=bx_j$ and $\tau=\sqrt{ab}t/m$.
	The EOM can then be rewritten as
	\begin{alignat}1
	\ddot Q_j&=e^{Q_{j-1}-Q_j}-e^{Q_j-Q_{j+1}},
	\label{EOM3}
	\end{alignat}
	where the time derivative is with respect to $\tau$.
	Define $e^{\frac{Q_j-Q_{j+1}}{2}}=2a_j$ and $\dot Q_j=2b_j$. 
	Then, the EOM of the Toda lattice under the periodic boundary condition \cin{become}
	\begin{alignat}1
	&\dot a_j=a_j(b_j-b_{j+1}),
	\nonumber\\
	&\dot b_j=2(a_{j-1}^2-a_j^2),
	\label{EOMp}
	\end{alignat}
	with $a_{j+q}=a_j$ and $b_{j+q}=b_j$.
	Note that $\prod_{j=1}^q a_j=2^{-q}$ holds by definition.
	Finally, the EOM (\ref{EOMp}) can be cast into the Lax equation \cite{Flaschka:1974aa}
	\begin{alignat}1
	\dot L(t)=[B(t),L(t)], 
	\label{Lax}
	\end{alignat}
	where the Lax pair $L$ and $B$ is defined by the action on the wave function $\varphi_j$
	with $\varphi_{j+q}=\varphi_j$ such that
	\begin{alignat}1
	&(L\varphi)_j=a_{j-1}\varphi_{j-1}+b_j\varphi_j+a_j\varphi_{j+1},
	\nonumber\\
	&(B\varphi)_j=a_{j-1}\varphi_{j-1}-a_j\varphi_{j+1}.
	\end{alignat}
	Instead of solving \cin{Eq.~(\ref{Lax})}, it is convenient to consider the eigenvalue problem for $L(t)$ \cite{Date:1976aa}:
	\begin{alignat}1
	L(t)\varphi_n(t)=\lambda_n\varphi_n(t).
	\label{LEig}
	\end{alignat}
	Along with the auxiliary conditions $\dot \varphi_n(t)=B(t)\varphi_n(t)$ and $\dot\lambda_n=0$,
	\cin{Eq.~(\ref{LEig})} is equivalent to the Lax equation (\ref{Lax}).
	
	\begin{figure}[h]
		\begin{center}
			\begin{tabular}{c}
				\includegraphics[width=0.9\linewidth]{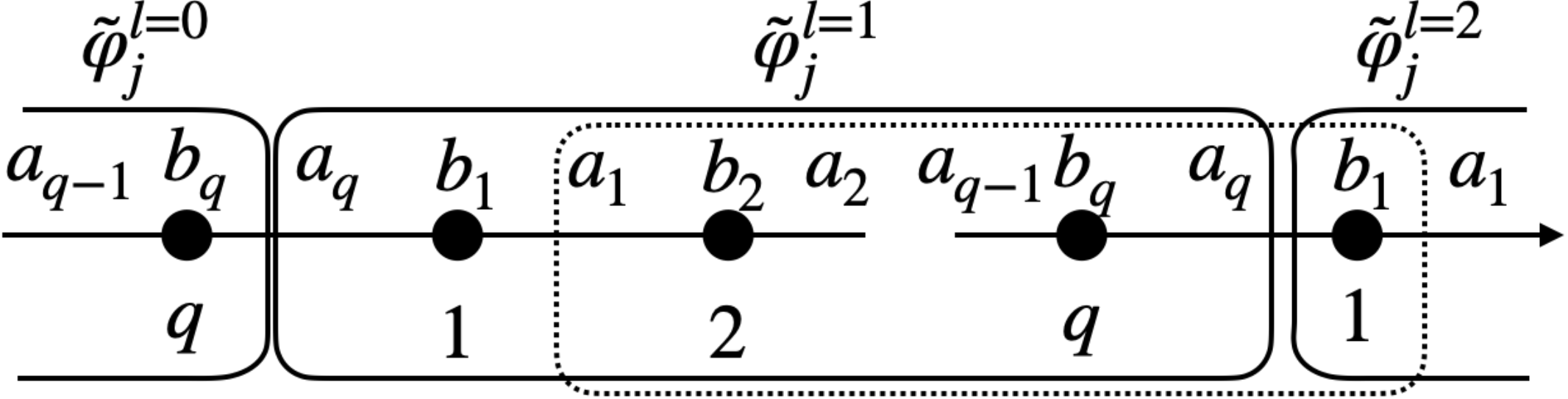}
			\end{tabular}
			\caption{
				The infinite chain of the periodic Toda lattice. The unit cell is composed of 
				$q$-sites labeled by $j=1,\dots,q$, on which $a_j$ and $b_j$ are set in the Lax operator.
				The unit cells are distinguished by the label $l$. The dashed square is another choice of the unit cell, starting with
				$j=2$ and ending with $1$.
				The dashed square is a unit cell for the computation of $b_1(t)$.
			}
			\label{f:il}
		\end{center}
	\end{figure}
	
	For the exact solution, 
	it is convenient to  consider an infinite system with the same $q$-periodicity \cite{Date:1976aa}. 
	\cin{To this end, let} us extend $j\rightarrow ql+j$ where $l$ takes integers while  $j=1,2,\cdots,q$, and 
	impose $a_{ql+j}=a_j$ and $b_{ql+j}=b_j$ as shown in Fig. \ref{f:il}. The wave function of $L$ is \cin{also} extended to 
	$\tilde\varphi_{ql+j}\equiv \tilde\varphi_j^l$, which is regarded as a $q$-vector $(\tilde\varphi^l)_j$ specified by $l$. 
	To write the corresponding Lax eigenvalue equation
	(\ref{LEig}), we introduce the matrix-valued operator 
	$\hat L(t)$ 
	\begin{alignat}1
	\hat L&=\left(
	\begin{array}{ccccccc}
	b_1&a_1&&&&& a_q\delta^{*q}\\
	a_1&b_2&a_2&&&&\\
	&a_2&b_3&a_3&&&\\
	&&a_3&&&&\\
	&&&&\ddots&a_{q-2}&\\
	&&&&a_{q-2}&b_{q-1}&a_{q-1}\\
	a_q\delta^q&&&&&a_{q-1}&b_q\\
	\end{array}
	\right),
	\label{LaxLop}
	\end{alignat}
	where $\delta=e^{\partial_i}$ and $\delta^*=e^{-\partial_i}$ are the forward and backward shift operators, respectively; $\delta f_i=f_{i+1}$
	and $\delta^* f_i=f_{i-1}$.
	For convenience, we further define 
	\begin{alignat}1
	{\cal K}=a_q\left(
	\begin{array}{ccccccc}
	&&&&&&1\\
	&&&&&\quad0&\\
	&&&&&&\\
	&&&\quad\iddots&&&\\
	&&&&&\\
	&&&&&&\\
	0&&&&&&\\
	\end{array}
	\right).
	\end{alignat}
	Then, the eigenvalue equation of the Lax operator is extended as: 
	\begin{alignat}1
	\hat L\tilde\varphi^l\equiv\left({\cal K}\Delta^{*}+{\cal V}+{\cal K}^T\Delta\right)\tilde{\varphi}^l=\lambda\tilde{\varphi}^l,
	\label{LEig_inf}
	\end{alignat}
	where 
	$\Delta\equiv\delta^q$ operates on $l$ such that $\Delta\tilde{\varphi}^l=\tilde{\varphi}^{l+1}$
	and likewise for $\Delta^*$, and ${\cal V}$ is the remaining part of $L$ acting on the same unit cell labeled by $l$.

	\begin{figure}[h]
		\begin{center}
			\begin{tabular}{cc}
				\includegraphics[width=0.5\linewidth]{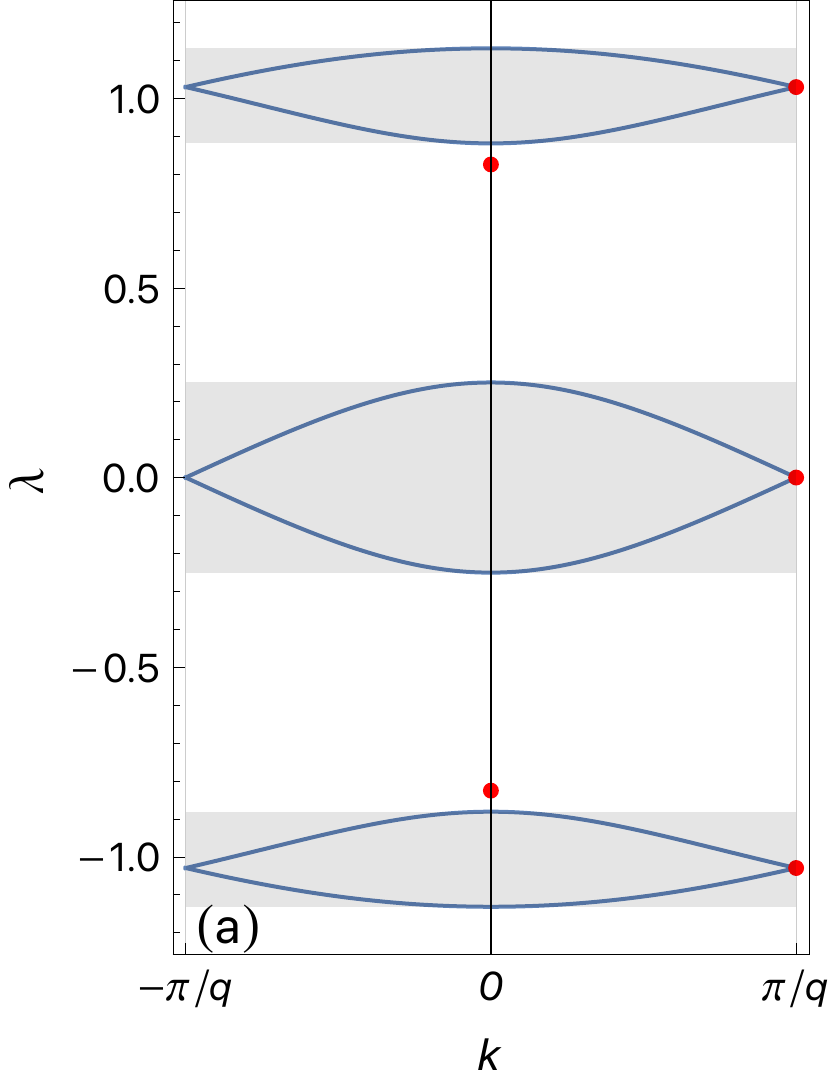}
				&
				\includegraphics[width=0.487\linewidth]{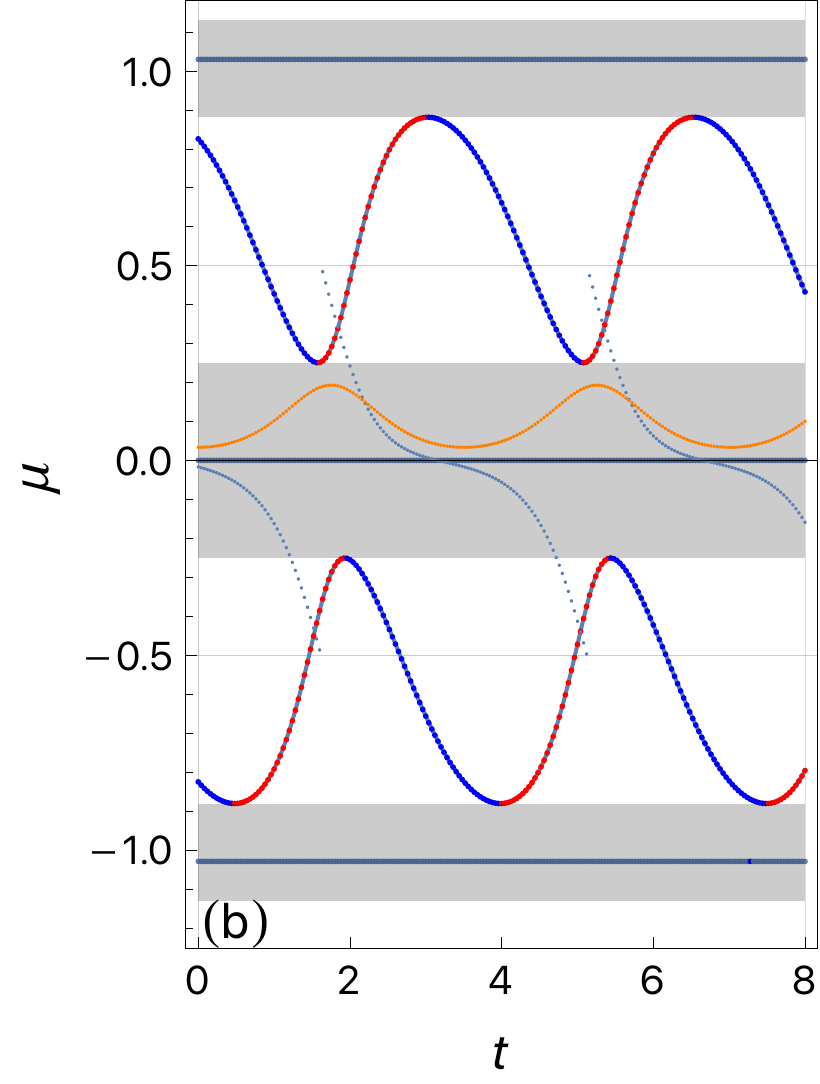}
			\end{tabular}
			\caption{(Color online)
				Example of the spectrum for the $q=6$ system with the initial condition \cin{$a_1(=a_4):a_2(=a_5):a_3(=a_6)=3:2:1$}, 
				while $b_j=0$.
				(a)
				The solid curves show $\lambda_n(k)$, and the red dots are the edge states at $t=0$.
				The shaded regions indicate the energies for which the Bloch states are allowed.
				(b)
				Time-evolution of the edge states in (a).
				The red and blue dots represent the edge states localized at the left and right ends, respectively, whereas
				the dark-blue dots forming straight lines are for the edge states within zero gaps that are prohibited to move.
				The shaded regions are the bulk bands.
				The small dark-blue and orange dots show the Berry phase $\Gamma_n$ (mod 1) of the lowest and middle bulk bands, respectively,
			}
			\label{f:lax_b}
		\end{center}
	\end{figure}

	First, we consider a bulk system with $-\infty<l<\infty$.
	The $q$-periodicity of finite systems is enhanced to $q$-translational invariance in \cin{Eq.~(\ref{LEig_inf})}.
	Then, it follows from the Bloch theorem that  $\tilde\varphi^l_j=\tilde\varphi_{ql+j}$ is 
	simultaneously an eigenstate of the translation operator $\Delta$ as well as of $\hat L$.
	Thus, we set 
	\begin{alignat}1
	\tilde\varphi^l_{j}=e^{ikql}\varphi_j(t,k),\quad (j=1,\dots,q)
	\label{Blo}
	\end{alignat}
	where the Bloch function $\varphi_j(t,k)$ satisfies $\varphi_{j+q}(t,k)=\varphi_j(t,k)$. 
	Note here that the $t$-dependence is explicitly written.
	Acting on the Bloch function, the Lax operator $\hat L(t)$ in (\ref{LaxLop}) becomes 
	the Bloch--Lax operator $L(t,k)$ in which the difference operators in $\hat L(t)$ are
	replaced by $\delta^q\rightarrow e^{iqk}$ and $\delta^{*q}\rightarrow e^{-iqk}$. 
	We note that \cin{$|k|\le\pi/q$}.
	In Fig. \ref{f:lax_b} (a), an example of the bulk spectrum is shown, which is the system of $q=6$ particles 
	with the initial condition $a_1=a_4,a_2=a_5,a_3=a_6\propto 3,2,1$. 
	The spectrum comprises six bands, of which three pairs are
	gapless, forming \cin{three bands}. This is because the minimum period is 3. 
	Therefore, this feature is also true for a more generic system with $q=3N$ particles if a similar initial condition 
	with period-3 is imposed. In this case, $N$ bulk bands appear in each shaded region in Fig. \ref{f:lax_b} (a)
	to form a single folded band. Thus, a total of three bands are separated by bulk gaps.
	
	Next, we consider \cin{Eq.~(\ref{LEig_inf})} under \cin{the open boundary condition}.
	The exact solutions of the Toda lattice under the periodic boundary condition were 
	obtained by using the edge states of $\hat L(t)$ \cite{Date:1976aa},
	where the edge states indicate the localized states at \cin{boundaries}, which are absent \cin{in} the bulk system.
	For numerical computations, a theory of edge states based on the Hermiticity of Hamiltonians, 
	developed for quantum tight-binding Hamiltonians, is convenient  \cite{Fukui:2020aa}. 
	Let $\tilde\varphi^l$ be defined on the semi-infinite line $l\ge1$. The unit cell at $l=1$ is referred to as the left end.
	Equation (\ref{LEig_inf}) is explicitly given by the following recurrence relation:
	${\cal K}\tilde\varphi^{l-1}+{\cal V}\tilde\varphi^l+{\cal K}^T\tilde\varphi^{l+1}=\lambda\tilde\varphi^l$.
	In particular, the initial equation for $l=1$ is given by:
	\begin{alignat}1
	&{\cal K}\tilde\varphi^0+{\cal V}\tilde\varphi^1+{\cal K}^T\tilde\varphi^2=\lambda\tilde\varphi^1.
	\label{Laxoeqx}
	\end{alignat}
	Thus, to define the system for $l\ge1$, 
	${\cal K}\tilde\varphi^0=0$ is imposed as the initial condition 
	of the recurrence \cin{relation.}
	Such vector $\tilde\varphi^0$ is generically given by 
	$(\tilde\varphi^{0})^T=\left(\chi^T,0\right)$,
	where $\chi$ is a $(q-1)$-vector. 
	Now, assume a generalized Bloch state given by $\tilde\varphi^l=e^{iqKl}\tilde\varphi^0$
	with a complex wavenumber $K=k+i\kappa$. Then, it is shown that  
	not only is \cin{Eq.~(\ref{LEig_inf})} satisfied for all $l\ge1$, but also the Hermiticity of $L(t,k)$ is 
	guaranteed at the boundary, as discussed in Ref. \cite{Fukui:2020aa}.
	To be specific, the equation to be solved is
	\begin{alignat}1
	&\hspace*{-1mm}\left(
	\begin{array}{ccccc|c}
	b_1&a_1&&&& e^{-iqK_n}a_q \!\!\!\!\\
	a_1&&&&&\\
	&&\ddots&&&\\
	&&&&a_{q-2}&\\
	&&&a_{q-2}&b_{q-1}&a_{q-1}\\
	\hline
	\!\!e^{iqK_n}a_q\!\!\!\!\!\!&&&&a_{q-1}&b_q\\
	\end{array}
	\right)\!\!\!
	\left(\begin{array}{c} \\\\  \\\!\! \chi_n \!\! \!\!\\ \\ \\ \hline 0 \end{array} \right)
	\!\!=\!
	\mu_n\!\!
	\left(\begin{array}{c}\\ \\  \\\!\! \chi_n\!\!\! \!\\ \\ \\ \hline 0 \end{array} \right),
	\label{La}
	\end{alignat} 
	where $\mu_n(t,K_n)$ denotes the energy of the edge states.
	Thus, \cin{it turns out that} the eigenvalues of the edge states
	are those of the $(q-1)\times(q-1)$ reduced matrix of the Bloch--Lax operator in the above equation, 
	and the last row gives the condition
	\begin{alignat}1
	e^{iqK_n}a_q\chi_{1n}+a_{q-1}\chi_{q-1,n}=0,
	\end{alignat}
	which determines \cin{$K_n=k_n+i\kappa_n$.}
	In the present case, $k_n$ is restricted to $k_n=0,\pi$, and $\kappa_n$ is a generic real number depending on $t$.
	For the wave function $\tilde\varphi_n^l$ to be localized at the left end,  $\kappa_n>0$ because
	$|\tilde\varphi_n^l|=e^{-q\kappa_n l}|\tilde\varphi^0|$.
	When $\kappa_n<0$, it cannot be an edge state at this end; instead, 
	a state such as this is localized at the right end, if the system with the right end is considered (see \cin{Ref.~\cite{Fukui:2020aa}}).
	Thus, for a positive (negative) $\kappa$, the edge state is localized at the left (right) end.

	We first examine how these edge states are embedded in the bulk band structure. 
	In Fig. \ref{f:lax_b} (a), the edge states at $t=0$ are shown by red dots.
	Among the five edge states in the system with $q=6$ particles, three are fixed at the band touching points. 
	Because the bulk spectrum is time independent, the energies of these edge states must also be.
	Thus, the other two edge states within the open gaps are responsible for the time evolution of the solution of the periodic Toda lattice. 
	More generically, for a system of $q=3N$ particles with the same period-3 initial condition, 
	there appear $3N-1$ edge states, among which $N-1$ states in each band are pinned at gapless
	points, and the remaining two edge states are located in the gaps.
	
	Next, we show in Fig. \ref{f:lax_b} (b) the spectrum of the edge states 
	as a function of $t$. 
	The energies of the two edge states oscillate, whereas those of the three edge states
	are constant.
	The desired solution of $b_q(t)$ is obtained by simply using the time evolution of the edge states, such that
	\cin{\cite{footnote1}}
	\begin{alignat}1
	b_q(t)=\sum_{j=1}^qb_j(0)-\sum_{n=1}^{q-1}\mu_{n}(t).
	\label{ExaSol}
	\end{alignat}
	Using the hyperelliptic abelian integrals and the solution to Jacobi's inversion problem for the above equation, 
	the exact solution can be written using the Riemann theta function \cite{Date:1976aa}.
	This implies that the spectrum of the edge states $\mu_n(t)$  is sufficient to obtain the exact
	solution of the periodic Toda lattice.
	Here, let us turn our attention to the wave function $\varphi_n(t,k)$, which
	provides further information about the end at which the edge states are localized.
	Such a feature is irrelevant to the exact solutions, but informs us of the topological property of the
	solution.
	From Fig. \ref{f:lax_b} (b), 
	we find that when $t=0$, the edge states are localized at the right end and when they touch the bulk bands
	under time evolution, 
	they become localized at the opposite left end. 
	Generically, the energies of the edge states oscillate as time evolves and when they touch the lower (upper) bulk band, 
	the right (left) edge states change to the left (right) edge states.
	This is one of the characteristic properties of topological edge states.
	
	\begin{figure}[h]
		\begin{center}
			\begin{tabular}{cc}
				\includegraphics[width=0.5\linewidth]{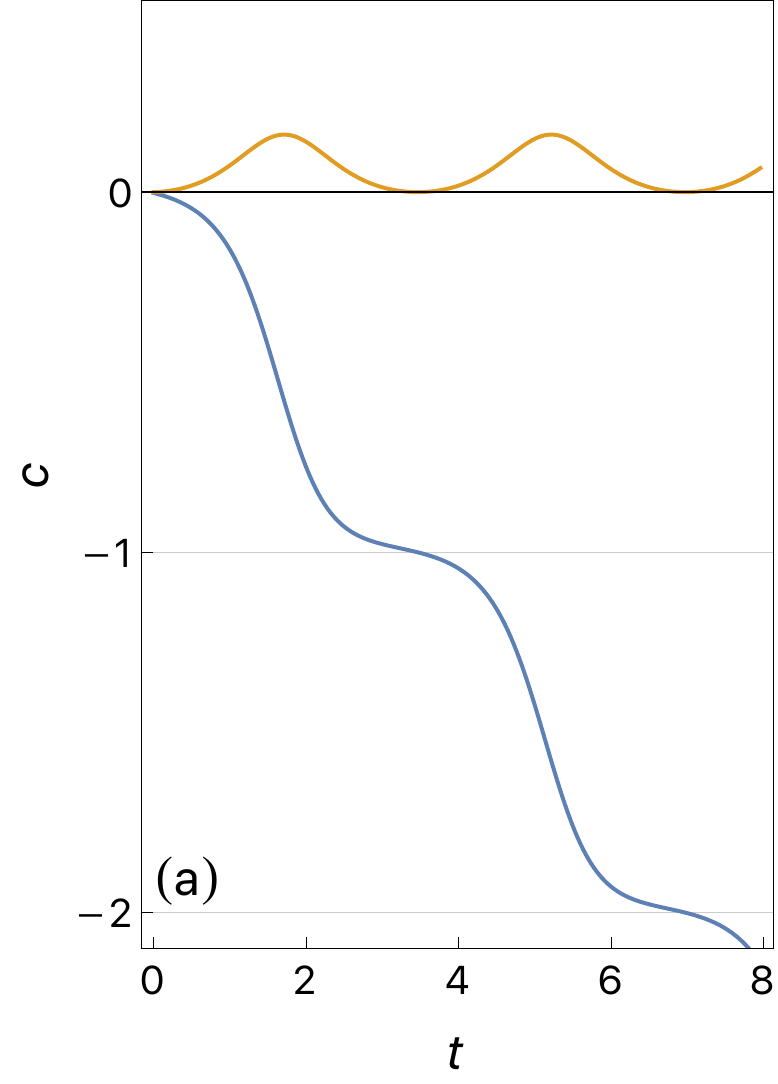}
				&
				\includegraphics[width=0.5\linewidth]{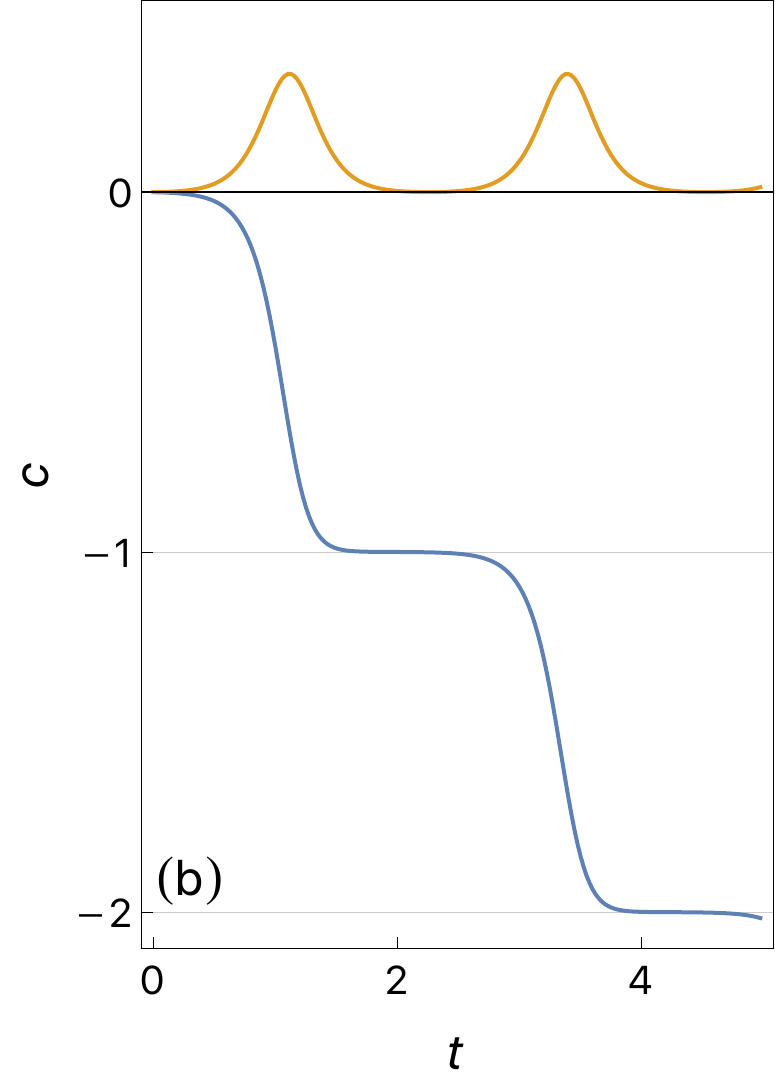}
			\end{tabular}
			\caption{(Color online)
				(a) Berry fluxes $c(t)$ of the lowest and middle (orange) band for the same system as 
				in Fig. \ref{f:lax_b}.
				(b) the same system but for the initial condition, $a_1:a_2:a_3=20:2:1$, 
				with a stronger nonlinearity than (a).
			}
			\label{f:chern}
		\end{center}
	\end{figure}
	
	This feature of the Lax operator $L(t,k)$  is reminiscent of 
	the Hamiltonian of the Thouless pump \cite{Thouless:1983fk}, of which the Hamiltonian is generally a function of $t$ and $k$,
	$H(t,k)$ with $H(t+T_0)=H(t,k)$ and $H(t,k+2\pi)=H(t,k)$.
	If the edge states of the Toda lattice have the same topological origin as the Thouless pump, 
	they are guaranteed by the bulk topological invariants, the Chern numbers.
	This motivated us to compute the Berry phases and Berry fluxes associated with 
	the \cin{Bloch} wave function \cin{$\varphi(t,k)$} in Eq. (\ref{Blo}).

	First, we introduce the Berry phase:
	\begin{alignat}1
	\Gamma_n(t)&=\frac{1}{2\pi i}
	\int_{-\pi}^\pi  \cin{dk}A_n(t,k),
	\end{alignat}
	where $A_n(t,k)$ is the Berry connection \cin{associated with the $n$-th band}, which is defined as
	\begin{alignat}1
	A_n(t,k)=\varphi_n^\dagger(t,k)\partial_k\varphi_n(t,k).
	\end{alignat}
	The Berry phase describes the charge polarization in quantum systems \cite{Marzari:2012aa}.
	In Fig. \ref{f:lax_b} (b), the Berry phases $\Gamma_n(t)$ for the lowest and middle bands \cin{are shown modulo 1}.
	\cin{We see that} the Berry phase of the lowest band has a nontrivial winding number. 
	Next, we introduce the Berry flux, defined as:
	\begin{alignat}1
	c_n(t)=\Gamma_n(t)-\Gamma_n(0).
	\end{alignat}
	This is equivalent to
	\begin{alignat}1
	c_n(t)&=\frac{1}{2\pi i}
	\int_0^t\!\!dt\!\!\int_{-\pi}^\pi \!\!\cin{dk F_{n}(t,k),}
	\end{alignat}
	where $F_n(t,k)$ is the Berry curvature, which is defined as
	\begin{alignat}1
	F_n(t,k)=\partial_t\varphi_n^\dagger(t,k)\partial_k\varphi_n(t,k)-{\rm c.c}.
	\end{alignat}
	For a periodic solution with a period $T_0$, $c_n(T_0)$ is guaranteed to be an integer referred to as the Chern number.
	
	\begin{figure}[h]
		\begin{center}
			\begin{tabular}{cc}
				\includegraphics[width=0.5\linewidth]{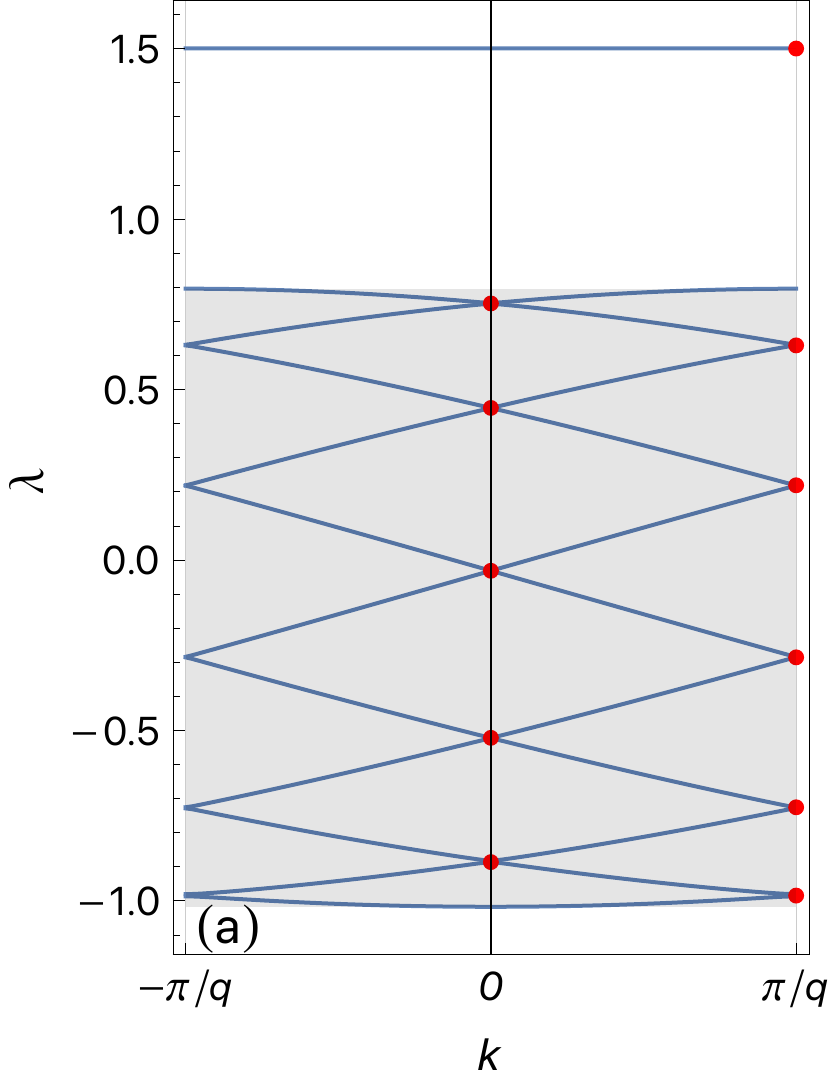}
				&
				\includegraphics[width=0.47\linewidth]{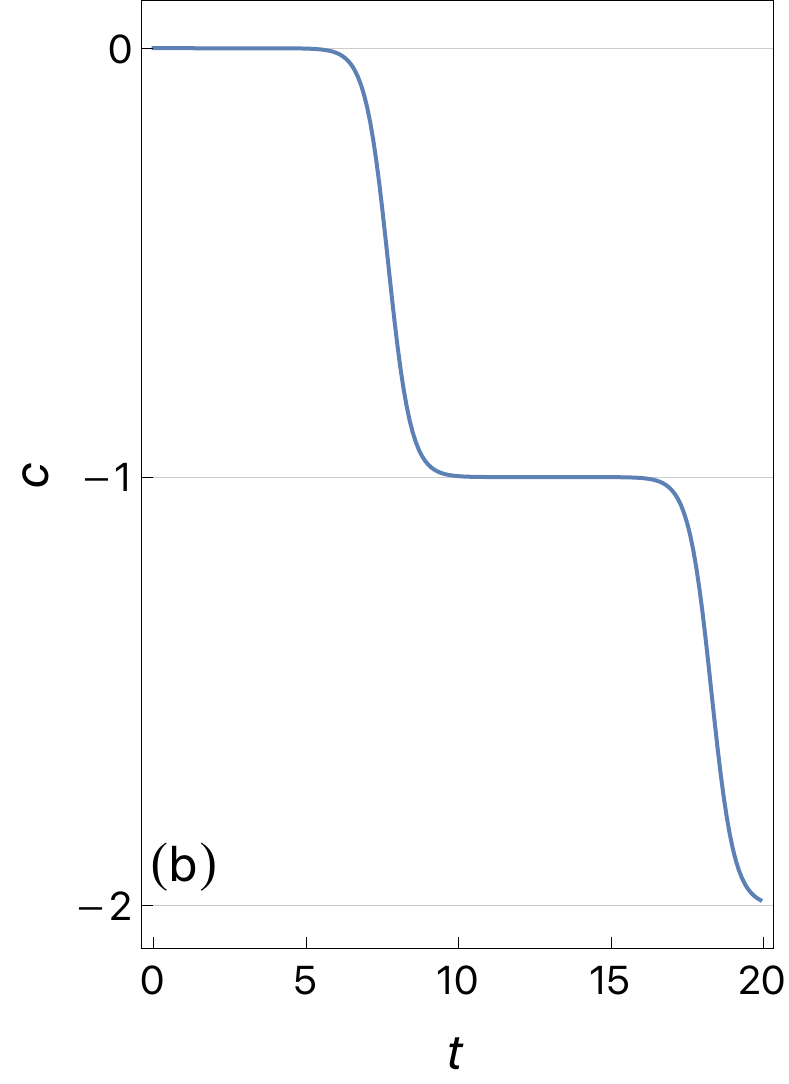}
			\end{tabular}
			\caption{(Color online)
				(a) Bulk spectrum of the system with $q=12$ for the exact cnoidal wave solution. 
				(b) Berry flux of (a).
			}
			\label{f:cnoidal}
		\end{center}
	\end{figure}
	
	In Fig. \ref{f:chern} (a), we show $c(t)$ for the system in Fig. \ref{f:lax_b} calculated by the method 
	developed in Ref. \cite{FHS05}. 
	We see that for just one period, $c$ changes by $-1$.
	This implies that the Chern number of the lowest band is $c=-1$. 
	This bulk topological invariant guarantees  
	the existence and stability of the edge states in the lower gap.
	By contrast, the second band does not have nonzero Chern numbers; hence, it is a topologically trivial band.
	Thus, the edge states in the upper gap are topologically stable, 
	because they are associated with the sum of the Chern numbers
	below the gap $-1+0=-1$ \cite{Hatsugai:1993aa,Hatsugai:1993fk}.
	For other multiband \cin{systems not} shown here, we also observe  similar behavior of the edge states in open gaps.  
	The stability of the edge states indicates the stability of the solutions of the Toda lattice; thus, we conclude that
	the solutions of the periodic Toda lattice are topologically stable.
	Remarkably, when substantial nonlinearity exists, as shown in Fig. \ref{f:chern} (b), 
	$c(t)$ plateaus at the integers $c=-n$.
	From the distance between these plateaus, we can easily determine the Chern numbers for one period of the solutions for (\ref{EOMp}).
	
	This feature is also true for the cnoidal wave of the periodic Toda lattice \cite{Toda:1967aa,Toda:1967ab}. 
	In Fig. \ref{f:cnoidal}, we show the bulk spectrum and
	Berry flux of the $q=12$ system for the exact single-mode cnoidal wave solution 
	with very large nonlinearity. 
	As expected, this is a two-band system \cite{Date:1976aa},  
	and the Berry flux exhibits very sharp plateaus, from which  
	we obtain the Chern number of $c=-1$ for one period.
	
	This observation raises the question of whether the Bloch--Lax operator $L(t,k)$ allows topologically trivial solutions, 
	such as for the Thouless pump.  
	In the Thouless pump, 
	not only the edge states but also the bulk states depend on time, such that at a certain time, gap-closing 
	may occur, resulting in a transition between the nontrivial and trivial phases \cite{Nakajima:2016aa,Lohse:2016aa}.
	However, in the present case, the bulk spectrum is time independent, implying that 
	such a point-like gap closing phenomenon is absent. 
	Therefore,  
	it is natural to expect the Lax operator to have no trivial phase; hence,  
	the cnoidal wave and more generic solutions of the periodic Toda lattice are topologically stable.
	
	In conclusion, we revealed the topological properties of the edge states that are used to obtain the
	exact solution of the periodic Toda lattice. We attributed the stability of nontopological solitons to the topological properties of the Lax operator.
	
	This work was supported in part by a Grant-in-Aid for Scientific Research 
	(22K03448) from the Japan Society for the Promotion of Science.


\begin{thebibliography}{21}
\bibitem{footnote1}{	For variables other than $b_q$, choose the appropriate unit cell; for example, that surrounded by the dashed square
		in Fig. \ref{f:il}, from which one can compute $b_1(t)$.}
\end{thebibliography}

\end{document}